\documentclass[a4paper]{article}

\usepackage{icrc2013}

\title{Constraining UHECR source models by the TA SD energy spectrum}

\shorttitle{ICRC 2013 Template}

\authors{
E. Kido$^{1}$,
O.E. Kalashev$^{2}$,
for the Telescope Array Collaboration.
}

\afiliations{
$^1$ Institute for Cosmic Ray Research University of Tokyo, Kashiwa, Chiba, Japan \\
$^2$ Institute for Nuclear Research of the Russian Academy of Sciences Moscow, Russia \\
}

\email{ekido@icrr.u-tokyo.ac.jp}

\abstract{
We have published the energy spectrum of ultra-high energy cosmic rays (UHECRs) for $E > 10^{18.2}$ eV 
from our first 5-year observation with the surface detectors of Telescope Array. 
We found two breaking points in the power-law spectrum, a flattening known as the "ankle" or "dip" at $E = 10^{18.66}$ eV, and a 
steepening at $E = 10^{19.72}$ eV. 
The spectral features must be related to distribution of cosmic ray sources, 
injection spectrum, and energy loss processes during propagation in inter-galactic space.
We constructed a model of source distribution of UHECRs and energy loss processes in the CMB photon 
field to calculate energy spectrum to be observed at the Earth. 
The model includes injection energy spectra at the sources, energy loss processes, and evolution of 
cosmic ray fluxes with red shift. 
We consider two cases that the UHECR sources are distributed uniformly, or same as galaxies in the local universe. 
We found that the spectrum can be well fitted with the model. 
We discuss to give a constraint to the model parameters, e.g. the injection power indices and cosmological evolution from the observed data.
}


\begin{document}
\maketitle

\section{Introduction}

The Telescope Array (TA) \cite{TA1} is a cosmic-ray observatory of the largest area in the northern hemisphere.
TA experiment has 507 surface particle detectors (SDs) with 1.2 km spacing covering about 700 km$^{2}$ ground area.
SDs are surrounded by 3 fluorescence detector (FD) stations which consist of 12 or 14 FDs.
The duty cycle of the SD array is greater than 95 \% throughout 5-year observation, whereas the FD duty cycle is about 10 \% because observation is possible only in moon-less clear night.
The TA SD energy spectrum given in Ref.\cite{Dmitri} is obtained by the 4-year observation.
This energy spectrum shows the ankle at 10$^{18.66}$ eV and the cutoff at 10$^{19.72}$ eV in 5.5 $\sigma$. 
Both of these features are firstly observed by High Resolution Fly's Eye (HiRes) \cite{HiResSpec} and also confirmed by Pierre Auger Observatory (PAO) \cite{PAOSpec}. 
We analyzed this energy spectrum with additional 1-year data.

Cosmic ray mass composition is quite important in interpreting cosmic ray spectral features. 
The result of the measurement of Xmax by TA FDs in stereo mode is consistent with proton models above $10^{18.2}$ eV \cite{Tameda}.
In other experiments, HiRes reported the consistency with proton models \cite{HiResXmax} but PAO reported the significant discrepancy between data and considered proton models \cite{PAOXmax}.

Several models have been proposed to explain the spectral features like the flattening around 10$^{18.7}$ eV or the steepening at 10$^{19.72}$. The dip model proposed by Berezinsky et al. \cite{dipmodel1} explains the ankle by the energy dependence of the $e^{+}$ $e^{-}$ pair production process when protons from extragalactic sources interact with cosmic microwave background (CMB) photons. The cutoff can be also explained by the interaction of protons with cosmic microwave background (CMB) photons. This process is known as GZK mechanism \cite{GZK1}.
Therefore both of the ankle and the cutoff feature are given only by protons from extragalactic sources with these processes.

In this paper, we compared the observed TA SD energy spectrum with this source model in detail. In the next section, the model calculations and the fitting procedure are explained, and the fitted results are shown in section \ref{results}. The discussion and conclusions are described in section \ref{discussion} and \ref{conclusion}.
\\
\section{Methodology}\label{method}

We firstly assumed uniform source distribution as cosmic ray sources.
In the latter part of this section, we also show the methodology when we assume local large scale structure (LSS) source distribution using galaxy distribution. This is a reasonable assumption if the astrophysical sources are UHECR sources. 
We also assumed that all the cosmic ray sources have the same power-law spectrum $\alpha E^{-\gamma}$, where $\alpha$ and $\gamma_g$ are the free parameters. The evolutions of source number densities are also applied uniformly as $(1 + z)^m$, where m is the free parameter.

The energy loss processes of UHECR protons in the CMB photon field is treated with the formulation in Ref.\cite{HKPT1} for energy spectrum below $2 \cdot 10^{19}$ eV. In this formulation the energy loss time of protons is given by a polynomial of $\log$ E. 
The redshift energy loss is also taken into account as $1/E \cdot dE/dz = 1/(1 + z)$. 
We employ CRPropa v2.0 \cite{CRPropa} for energy spectrum with energies greater than $2 \cdot 10^{19}$ eV 
to simulate energy losses to consider the distribution of energy losses in each interaction.
\\
\\
\subsection{Uniform source distribution}\label{uniform}

Using an uniform source distribution and the energy loss processes, the expected energy spectrum is calculated with the following procedure that is similar to the section 2.3 of Ref.\cite{HKPT2} for energy spectrum below $2 \cdot 10^{19}$ eV.
Firstly we start from the energy spectrum of a single source.
We assume that each source isotropically emits cosmic rays.
The expected single source integral energy flux on the earth $F_{exp}^{1}(E)$ at a proper distance D can be expressed by 
\begin{equation}
F_{exp}^{1}(E) = \frac{J^0(E_0)}{4 \pi D^2 (1 + z)},
\end{equation}
where $J^0(E_0)$ is the source power above $E_0$ at the source and $E_0$ is the energy threshold at the source. 
The proper distance $D(z)$ is given by 
\begin{equation}
D(z) = \int_0^{z} \frac{c dz'}{H(z')}, 
\end{equation}
where c is the speed of light and $H(z)$ is the Hubble parameter at the redshift $z$.

The total integral flux $F_{exp}(E)$ within the redshift $z_{max}$ can be described by
\begin{equation}
F_{exp}(E) = \frac{c n_{src}(z = 0)}{4 \pi} \int^{z_{max}}_{0} dz \frac{J^0(E_0) (1 + z)^{m - 1}}{H(z)},
\end{equation}
where $n_{src} (z = 0)$ is the UHECR source density at $z = 0$ and $m$ is the assumed source number evolution. 
From the assumption, $J^0(E)$ can be given by 
\begin{equation}
J^0(E) = \int^{E_{max}}_{E} \alpha E'^{- \gamma_g} dE',
\end{equation}
where $E_{max}$ is the acceleration limit of cosmic rays at the source.
$E_0$ can be calculated by the numerical integration of $dE/dz$ from the energy $E$ with the energy loss processes of protons which are explained above. 
$F_{exp} (E)$ is actually calculated using these calculation processes when $z_{max} = 5$, $H(z) = H_0 \sqrt{(1 + z)^3 \Omega_m + \Omega_{\Lambda}}$,  $H_0 = 72$ km/s/Mpc, $\Omega_m = 0.27$, $\Omega_{\Lambda} = 0.73$ and $E_{max} = 10^{21}$ eV.
The differential energy flux $f_{exp} (E) = dF_{exp} (E)/dE$ is also numerically calculated with given $F_{exp} (E)$.

Monte Carlo simulations of the trajectories of protons are carried out using CRPropa v2.0 to calculate energy spectrum above $2 \cdot 10^{19}$ eV.
The uniform source distribution is set from 0 Mpc to 1400 Mpc, which is well beyond the GZK horizon. 
More than about $10^5$ trajectories are obtained and we calculate differential energy flux from them.
We connect this result to the flux from lower energies at $2 \cdot 10^{19}$ eV to obtain the final energy spectrum.
We calculated expected number of events $N_{exp}$ in each energy bin from this calculated differential energy flux.
In this calculation, we used the same acceptance and energy resolutions as in Ref.\cite{Dmitri}.
We compared $N_{exp}$ with the data.

The model fitting of the data is carried out for energies above $10^{18.2}$ eV. 
The best-fit four model parameters $\alpha$, $\gamma_g$, $m$ and the energy scale are determined by the fitting.
The best-fit $\alpha$ and the energy scale are also determined for the fixed $\gamma_g$ and $m$ in order to determine the confidence region of $\gamma_g$ and $m$.

The fitting procedures are the following.
Here, we employed a binned likelihood analysis method considering Poisson probability distribution of number of events in each energy bin.
We define the likelihood function
\begin{equation}
L = \prod_{i=1}^{N} P(N_i;N_{exp}(E_i)),
\end{equation}
where $P(n;\mu) = \mu^{n} \exp(-\mu)/n!$, $N_i$ is the observed number of events in the i-th energy bin, $E_i$ is the i-th energy and N is the number of bins. In this case, $N = 21$.
The likelihood ratio can be given by $\lambda = \prod_i P(N_i;N_{exp}(E_i))/\prod_i P(N_i;N_i)$, and
the quantity $- 2 \ln \lambda$ follows $\chi^2$ distribution.
The degree of freedom of the $\chi^2$ corresponds to $N$ minus number of fitting parameters.
The best-fit parameters are determined by minimizing this $\chi^2$.
Here, we define p-values as the probability of obtaining larger $\chi^2$ minimum than the observed one.
We calculated p-values for the fixed parameters to draw the confidence region of the parameters.
\\

\subsection{LSS source distribution}\label{LSS}

As a model of UHECR sources along the large scale structure, we used about 110,000 galaxy samples of 2 Mass Extended Source Catalogue (XSCz)\cite{XSCz}.
We selected the galaxies in the XSCz catalog with distances smaller than 250 Mpc and apparent
magnitudes brighter than 12.5 in the Ks band (2.2 $\mu$m).
We assumed an uniform matter distribution beyond 250 Mpc.

The galaxies within 5 Mpc were not included because they do not represent a proper statistical sample of LSS. 
We introduced a weighting factor for each selected galaxy to take into account faint galaxies below the limit 12.5 with the distribution of absolute magnitudes of galaxies as proposed in Ref.\cite{TAAniso}.

We calculated the distance dependence of $\sum_{i} w_i A_i$ from the galaxies, where $w_i$ is the weight of each galaxy and $A_i$ is the relative TA SD exposure in the direction of the galaxy.
Then we simulated the propagation from the source distribution $\sum_{i} w_i A_i$ in the same way as Section.\ref{uniform}.

\begin{figure}[t]
      \centering
      \includegraphics[width=0.5\textwidth]{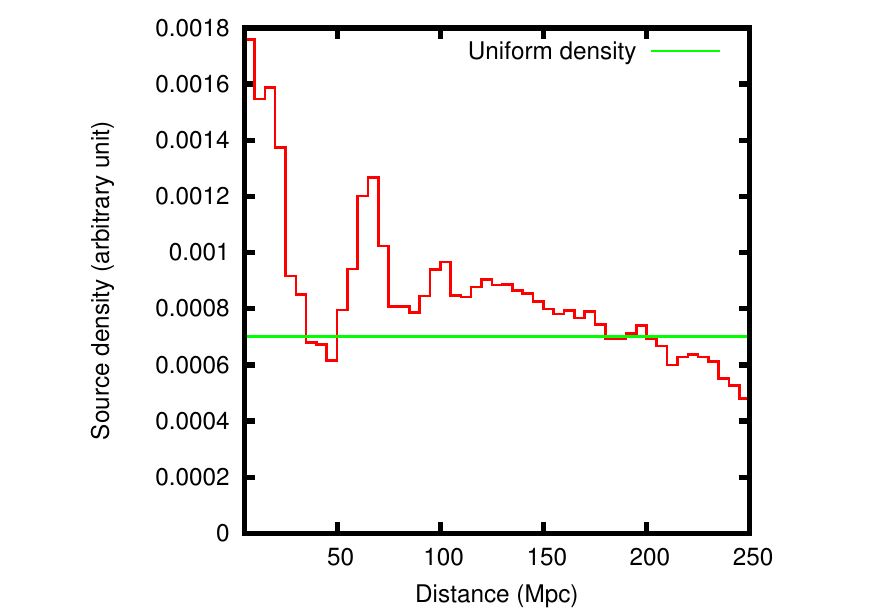}
      \caption{Calculated $\sum_{i} w_i A_i$ distribution of galaxies in the XSCz catalogue, where $w_i$ is the weight of i-th galaxy and $A_i$ is the relative TA SD exposure in the direction of the galaxy. In this figure, calculated $\sum_{i} w_i A_i$ within a slice of 5 Mpc divided by squared distance is presented.}
      \label{LSS_z}
\end{figure}

Fig.\ref{LSS_z} shows the calculated distance dependence of $\sum_{i} w_i A_i$ from the galaxies.

\section{Results}\label{results}

\subsection{Uniform source distribution}\label{uniform_result}

\begin{figure}[t]
      \centering
      \includegraphics[width=0.4\textwidth]{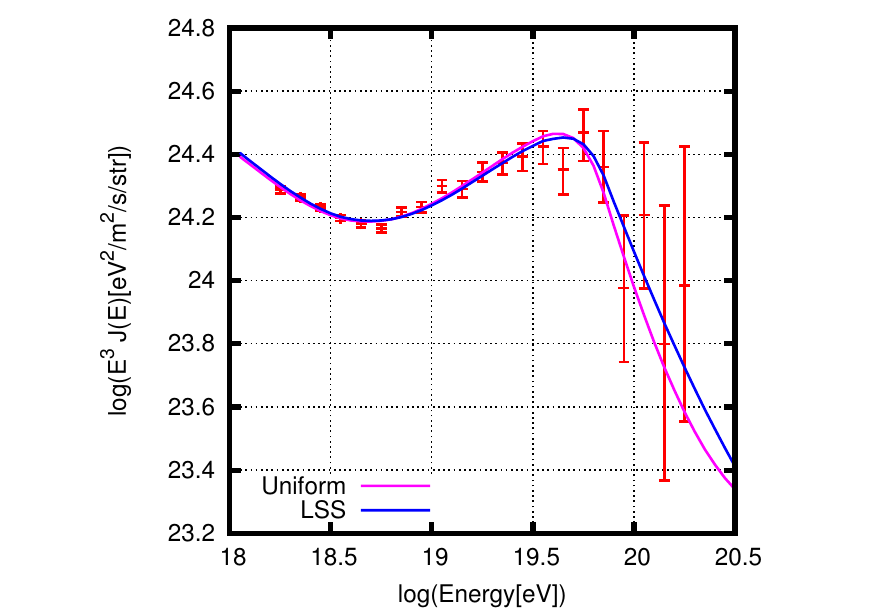}
      \caption{The data points of the observed energy spectrum by TA SD are drawn as red points. The best fit expected energy spectrum is drawn as a pink line when $\gamma_g = 2.36$,  $m = 4.5$ and UHECRs are uniformly distributed. $\chi^2$/d.o.f. is $20.2/17$. The best fit expected energy spectrum is drawn as a blue line when $\gamma_g = 2.39$,  $m = 4.4$ and UHECRs are distributed along the LSS. In this case $\chi^2$/d.o.f. is $18.5/17$.}
      \label{Espec_uni}
\end{figure}

We show the expected energy spectrum with source parameters $\gamma_g = 2.36$ and $m = 4.5$ determined by fitting the TA SD spectrum in Fig.\ref{Espec_uni}.
The expected energy spectra are calculated when $\gamma_g = 2.00, 2.01, \cdots 2.69$ and $m = 0.00, 0.01, \cdots 9.99$, and compared with the data in the same way.

\begin{figure}[t]
      \centering
      \includegraphics[width=0.5\textwidth]{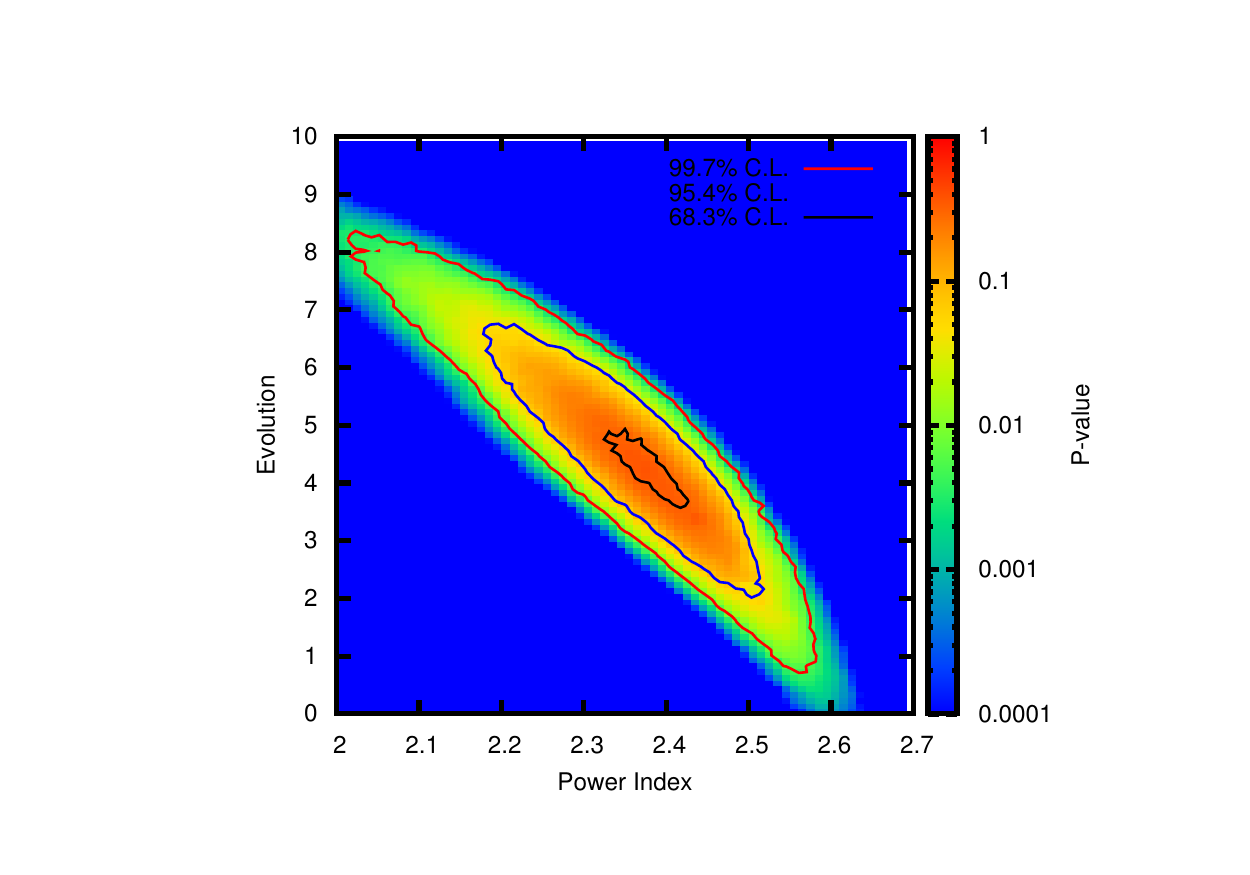}
      \caption{ P-values resulted from the values of likelihood ratios are plotted with a color map when UHECRs are uniformly distributed. The x-axis is the injection power index $\gamma_g$ of sources. The y-axis is $m$ of the number evolution $(1 + z)^m$ of sources.}
      \label{parameter_map}
\end{figure}

The p-values calculated from the $\chi^2$ are shown in Fig.\ref{parameter_map}.
The two parameters $\gamma_g$ and $m$ are correlated,
however we can give a constraint band to the parameters with the current statistics of TA as shown in Fig.\ref{parameter_map}.
$\gamma_g = 2.36^{+ 0.08}_{- 0.04}$ and  $m = 4.5^{+ 0.6}_{- 1.1}$ are obtained as a conclusion.

Note that the number evolution $(1 + z)^m$ are known for several classes of astrophysical objects, e.g. $m = 3$ for QSOs and $z < 1.3$ \cite{QSO_evol}, $m = 4.8$ for GRBs and $z < 1$ \cite{GRB_evol} and $m = 5$ for AGNs and $z < 1.7$ \cite{AGN_evol}. \\

\begin{figure}[t]
      \centering
      \includegraphics[width=0.5\textwidth]{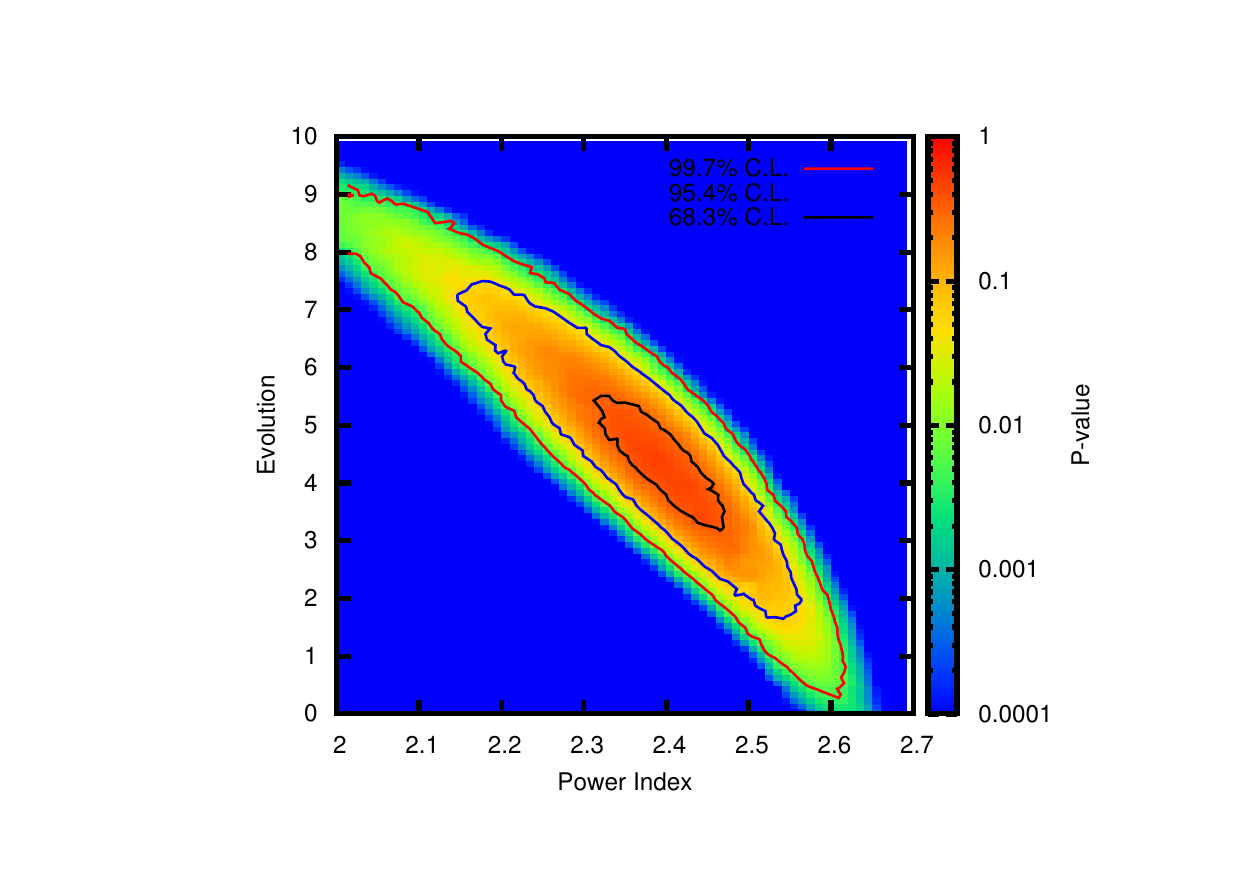}
      \caption{P-values resulted from the values of likelihood ratios are plotted with a color map when UHECRs are distributed along the LSS. The x-axis and y-axis are same as in Fig.\ref{parameter_map}.}
      \label{parameter_contour}
\end{figure}

\subsection{LSS source distribution}\label{LSS_result}

The blue line in Fig.\ref{Espec_uni} is calculated in the same procedure as the pink line except for the source distribution. 
In LSS source distribution, there are some regions locally over-dense because of local galaxy clusters like Virgo cluster as you can see in Fig.\ref{LSS_z}. 
Therefore the flux at the high-energy end expected from the LSS source distribution is larger than that from the uniform sources as seen in Fig.\ref{Espec_uni}.
The $\chi^2$ of the LSS source distribution in Fig.\ref{Espec_uni} is not much different from that of the uniform sources as a result because of large error bars of the data at the high energy end. 

The p-values calculated from the $\chi^2$ are shown in Fig.\ref{parameter_contour}.
$\gamma_g = 2.39^{+ 0.08}_{- 0.08}$ and  $m = 4.4^{+ 0.9}_{- 1.3}$ are obtained as a conclusion.

\section{Discussion}\label{discussion} 

The systematic uncertainty of TA SD energies is 22\% \cite{Dmitri}. 
This only changes the energy scale and does not affect the shape of the energy spectrum.
We included the shift of the energy scale in the fitting as a free parameter as described above.
Here, we define the shift of the energy scale as $\Delta \log E = \log E' - \log E$, where $E'$ is the shifted energy scale.
The energy scale $\Delta \log E = -0.01^{+0.02}_{-0.02}$ is obtained by the fitting if we assume uniform source distribution.
The energy scale $\Delta \log E = 0.02^{+0.04}_{-0.05}$ is determined by the fitting if we assume LSS source distribution.
So the observed energy scale is consistent with these models.
The systematic uncertainty of the energy scale is larger than the statistical fluctuation of that.
So this can have a big influence on the conclusions in this paper.

In case of strong evolution models, the larger component of energy flux is produced from the distant sources.
This situation can be strongly restricted from the observation of the diffuse gamma ray background.
We calculated secondary photons in the propagation of cosmic ray protons with CRPropa v2.0.
The sources with redshift below 0.7 are taken into account.
This redshift corresponds to the event horizon of $10^{18.2}$ eV cosmic ray protons. 
If $m > 7$, the energy spectrum of secondary photons exceeds the data points of the diffuse gamma ray background that is observed by Fermi Large Area Telescope\cite{Fermi}.

If the source density has the dependence of the maximum acceleration energy $E_{max}$ or luminosity dependence, our conclusions can be different \cite{Semikoz} \cite{Aloisio}. If the source density $n_{src}(E_{max}) \propto E_{max}^{-\delta}$, $\delta = 1.7$ and $\gamma_g = 2.0$, the energy spectrum becomes similar to the source with $\gamma_g = 2.7$ and $\delta = 0$ when we do not consider the evolution.
If the clustering in the arrival directions is observed in the future, this dependence can be discussed experimentally.

\section{Conclusions}\label{conclusion}

We calculated the expectation of the cosmic ray energy spectrum of uniform source distribution and LSS source distribution with pure proton model.
We fitted the TA SD spectrum with this expectation.
$\chi^2/d.o.f. = 20.2/17$ is obtained if we assume uniform source distribution.
$\chi^2/d.o.f. = 18.5/17$ is determined if we assume LSS source distribution.
Both source distributions can explain the data.

We investigated a compatibility of the TA SD spectrum and the expectation with source spectral index $\gamma_g$ and the evolution $m$ as free parameters.
We found that $\gamma_g = 2.36^{+ 0.08}_{- 0.04}$ and  $m = 4.5^{+ 0.6}_{- 1.1}$ are obtained in case of uniform source distribution.
$\gamma_g = 2.39^{+ 0.08}_{- 0.08}$ and  $m = 4.4^{+ 0.9}_{- 1.3}$ are determined if we assume LSS source distribution.
No evolution model is ruled out.

This result is determined mainly from the dip feature of the energy spectrum around $10^{18.7}$ eV and TA SD needs more statistics to discuss the compatibility at the high energy end which is above $10^{19.7}$ eV. We need more statistics in higher energies to rule out the uniform source distribution model.

\vspace*{0.5cm}
\footnotesize{{\bf Acknowledgment:}{The Telescope Array experiment is supported by the
Japan Society for the Promotion of Science through
Grants-in-Aid for Scientific Research on Specially Pro-
moted Research (21000002) “Extreme Phenomena in the
Universe Explored by Highest Energy Cosmic Rays”
and for Scientific Research (S) (19104006), and the
Inter-University Research Program of the Institute for
Cosmic Ray Research; by the U.S. National Science
Foundation awards PHY-0307098, PHY-0601915, PHY-
0703893, PHY-0758342, PHY-0848320, PHY-1069280,
and PHY-1069286 (Utah) and PHY-0649681 (Rutgers);
by the National Research Foundation of Korea (2006-
0050031, 2007-0056005, 2007-0093860, 2010-0011378,
2010-0028071, 2011-0002617, R32-10130); by the Rus-
sian Academy of Sciences, RFBR grants 10-02-01406a
and 11-02-01528a (INR), IISN project No. 4.4509.10
and Belgian Science Policy under IUAP VI/11 (ULB);
by the Grant-in-Aid for the Scientific Research (S) No.
19104006 by the Japan Society for the Promotion of Sci-
ence. The foundations of Dr. Ezekiel R. and EdnaWattis
Dumke, Willard L. Eccles and the George S. and Dolores
Dore Eccles all helped with generous donations. The
State of Utah supported the project through its Eco-
nomic Development Board, and the University of Utah
through the Office of the Vice President for Research.
The experimental site became available through the co-
operation of the Utah School and Institutional Trust
Lands Administration (SITLA), U.S. Bureau of Land
Management and the U.S. Air Force. We also wish to
thank the people and the officials of Millard County,
Utah, for their steadfast and warm support. We grate-
fully acknowledge the contributions from the technical
staffs of our home institutions and the University of Utah
Center for High Performance Computing (CHPC).}}

\end{document}